# Magnetism of noncolinear amorphous DyCo$_3$ and TbCo$_3$ thin films

Zexiang Hu[1], Ajay Jha[1], Katarzyna Siewierska[2], Ross Smith[1], Karsten Rode[1], Plamen Stamenov[1], J. M. D. Coey[1]*.

[1] School of Physics and Centre for Research on Adaptive Nanostructures and Nanomaterials, Trinity College Dublin, Ireland.
[2] Institute for Methods and Instrumentation for Synchrotron Radiation research (PS-ISRR), Helmholtz-Zentrum Berlin für Materialen und Energie GmbH (HZB), Albert-Einstein-Str. 15, 12489 Berlin, Germany

Abstract

Thw magnetization of sputtered films of amorphous Dy$_x$Co$_{1-x}$ and Tb$_x$Co$_{1-x}$ with x ≈ 0.25 is studied by SQUID magnetometry, anomalous Hall effect (AHE) and magneto-optic Kerr effect (MOKE) with the aim of understanding the temperature-dependent magnetic structure of the films. A square magnetic hysteresis loop with perpendicular magnetic anisotropy and coercivity that reaches 3.5 T in the vicinity of the compensation temperature is observed in films that are 10 - 20 nm thick. An additional anhysteretic soft component, seen in the magnetization measurements of some films, but not in their Hall or Kerr loops, is an artefact arising from material sputter-deposited on the sides of the substrate. The temperature-dependence of the net rare earth moment in the range 4 – 300 K is deduced, using the cobalt moment measured in amorphous Y$_x$Co$_{1-x}$. The single-ion anisotropy energy of the quadrupole moments of 4$f$ rare-earth atoms in the randomly-oriented local electrostatic field gradient exceeds the magnitude to their exchange coupling with the cobalt subnetwork, resulting in a sperimagnetic ground state where the spins of the noncollinear rare-earth subnetwork are modelled by a distribution of rare earth moments within a cone whose axis is antiparallel to the ferromagnetic axis **z** of the cobalt subnetwork. The reduced magnetization $\langle J_{-z}\rangle/J$ at $T$ = 0 is calculated from a single-atom Hamiltonian as a function of $\alpha$, the ratio of anisotropy to exchange energy per rare-earth atom for a range of angles $\theta$ between the local anisotropy axis and -**z** and then averaged over all directions in a hemisphere. The experimental and calculated values of $\langle J_{-z}\rangle/J$ are close to 0.7 at low temperature for both Dy and Tb. On increasing temperature, the magnitude of the rare earth moment and the local random anisotropy that creates the cone are reduced; the cone angle closes and the structure approaches collinear ferrimagnetism well above ambient temperature. An asymmetric spin flop of the exchange-coupled subnetworks appears in the vicinity of the magnetization compensation temperatures of 175 K for amorphous Dy$_{0.25}$Co$_{0.75}$ and 200 K for amorphous Tb$_{0.25}$Co$_{0.75}$. Partial single-pulse all-optical switching is investigated by X-ray photoelectron microscopy as a function of temperature in a Dy$_{0.25}$Co$_{0.75}$ film.

*jcoey@tcd.ie

14-12-23

# 1. Introduction

Interest in the magnetism of amorphous alloys of the rare earths with iron or cobalt dates from work in the 1970s and 1980s on the crystalline rare-earth intermetallic compounds, which were studied in an effort to understand the relevant exchange and crystal-field interactions [1,2], That research, which led to the discovery of $Nd_2Fe_{14}B$ [3], was spurred by a demand for cobalt-free permanent magnets at a time in the aftermath of the 1976 cobalt crisis when the best permanent magnets, Sm-Co and Alnico, were Co-based. Metal evaporation and sputtering techniques were then being perfected to produce thin films of the amorphous counterparts. A wealth of experimental information was acquired on the magnetism of amorphous rare-earth iron and rare-earth cobalt thin films [4] and a schematic classification of collinear and noncollinear magnetic structures in these amorphous alloys was developed [5]. A practical motivation was to develop perpendicular media for magneto-optical data recording [6] based on Curie point or compensation point writing. Amorphous ferrimagnetic $Gd_{0.25}Fe_{0.656}Co_{0.094}$ films were optimized in this context [7]. Later they were shown to exhibit helicity-dependent all-optical switching [8] and in 2012 repetitive single-pulse all-optical toggle switching (SPAOS) was discovered — a fast thermal response of the thin film to irradiation with sub-picosecond laser pulses [9]. The same effect is observed in binary amorphous a-$Gd_{0.30}Co_{0.70}$ films [10] but not in a-$Dy_{0.25}Co_{0.75}$ or a-$Tb_{0.25}Co_{0.75}$ films, where partial single-pulse all-optical switching is observed only for the first few pulses [11]. Complete one-shot switching of a-$Tb_{0.24}Co_{0.76}$ was only possible in microdots with microantennae [12] or in a 1.5 T bias field [13]. In this paper 'a-' is used to denote an amorphous alloy.

The amorphous alloys of Co with a heavy rare earth Gd, Tb, Dy, Ho, Er or Tm exhibit a ferrimagnetic compensation point $T_{comp}$ below or close to room temperature when the ratio of rare-earth R to Co is approximately 1:3 [4, 7, 14 - 17]; the compensation can be tuned by varying the rare earth content $x$ in a-$R_xCo_{1-x}$. The Gd alloys are collinear ferrimagnets (7,14), but Dy atoms in a-$DyCo_3$ were shown by $^{161}$Dy Mossbauer spectroscopy to adopt a noncollinear structure at low temperature [17] related to the random local electrostatic field acting on the electric quadrupole moment of the incomplete Dy 4f shell, which produces random local magnetic anisotropy. The resulting noncollinear ferrimagnetism, known as sperimagnetism, was modelled by a uniform distribution of rare earth moment orientations within a cone of half-angle $\theta_0$ aligned antiparallel to the cobalt magnetization direction (Fig 1). Alloys of Co with Y, a nonmagnetic rare earth, are soft ferromagnets with only a small induced moment on Y.

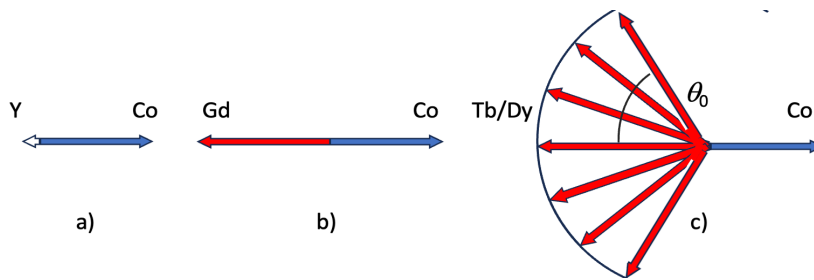

Figure 1. Schematic magnetic structures of amorphous alloys: a) ferromagnetic a-YCo$_3$ b) ferrimagnetic a-GdCo$_3$ and sperimagnetic a-DyCo$_3$ or a-TbCo$_3$.

Recent studies of a-R$_x$Co$_{1-x}$ films, R = Gd, Tb, Dy, relevant to the present work include investigations of the sperimagnetism [18-20], determinations of the atomic moments by X-ray magnetic circular and linear dichroism [21,22], reports of a bulk Dzyaloshinskii-Moriya interaction in a-GdFeCo [23] and a skyrmion lattice in DyCo$_3$ [24, 25]. More generally, there has been a recent revival of interest in these materials in the context of ferrimagnetic spintronics [26], where ultra-fast longitudinal spin dynamics in the picosecond range underpins all-optical switching, and large anisotropy fields in the vicinity of the compensation temperature facilitate fast transverse spin dynamics and high-speed domain-wall propagation. The amorphous structure has no centre of symmetry, and the amorphous rare earth – transition metal films have potential for spin-orbit torque switching.

As a preliminary to reinvestigation of the magnetism of the amorphous R-Co alloys with a heavy rare earth, we examined the magnetism of the a-Y$_x$Co$_{1-x}$ series [27], where yttrium serves as a non-magnetic proxy for the rare earths with 4f electrons. Lanthanum can also be used [18]. Films with a wide range of composition $0 < x \leq 0.55$ covered the appearance of magnetism at $x_c = 0.50$. We focussed on compositions with $x \approx 0.25$, an 1:3 Y:Co ratio, where the volumes occupied by Y and Co atoms in the structure are roughly equal, and the nearest neighbour coordination numbers were obtained by modelling the binary amorphous structure. The Co has a large orbital moment of 0.31 $\mu_B$ and a random local anisotropy of 5.2 K/atom (3.2 MJm$^{-3}$) but the alloy is nevertheless a soft ferromagnet because of exchange averaging in the Co exchange field [27]. The Co-Co exchange is actually greater in the amorphous state than in the crystalline state, and the Curie temperature of pure amorphous cobalt (if it existed) would be ≈ 2000 K [28,29]. The ratio of exchange energy to local Co anisotropy energy of a Co atom in a-YCo$_3$ is 220. However, we will show that exchange averaging is ineffective for Dy or Tb, because of the much stronger local random anisotropy on 4f atoms and the much weaker exchange field acting on the rare earth, which therefore tends to follow the local anisotropy direction, leading to a random noncollinear structure.

## 2. Thin Films

Films of a-R$_x$Co$_{1-x}$ with R = Dy or Tb and $x = 0.25$ were grown on oxidized silicon wafers by DC sputtering from separate 50 mm rare earth and cobalt targets in a Shamrock sputtering system with a base pressure of about $8 \times 10^{-8}$ torr. No metallic Pt or Ta underlayer or overlayer was used since we are interested in magnetic and transport properties of the amorphous material. Films were capped with a 2–4 nm thick layer of SiO$_2$ or Al$_2$O$_3$ to protect them from oxidation. Tb and Dy are chosen because they are the heavy rare earths with the largest negative quadrupole moments [30], which are necessary for local easy axis anisotropy in the amorphous R-Co atomic environment. Film composition was chosen to favour perpendicular anisotropy and ensure that compensation occurs below room temperature where the ferromagnetism of the cobalt subnetwork is practically temperature-independent. The concentration $x$ was determined from the calibrated deposition rates from the individual targets.

The films were characterized by X-ray diffraction and small-angle X-ray scattering using a Philips Panalytical X'pert Pro diffractometer [27]. No Bragg reflections from any R–Co intermetallic phase were seen in the diffraction patterns. The density and roughness of the film and cap layers as well as their thickness were fitted using the X'pert Reflectivity Pro software. Data on the materials of most interest here are listed in Table 1, including two a-$Y_{0.25}Co_{0.75}$ films for comparison [27].

One difference between our a-$Dy_xCo_{1-x}$ and a-$Tb_xCo_{1-x}$ samples is that the Dy films were stable in ambient conditions for a year of more, whereas the Tb films began to deteriorate after about a month. This difference may related to the different oxides associated with the two rare-earths. Dy forms a sesquioxide like most other trivalent lanthanides, but Tb forms a higher oxide '$Tb_4O_7$' which is a mixture of phases containing trivalent $4f^8$ and quadrivalent $4f^7$ terbium ions [31].

Table 1. Structural and magnetic data on amorphous thin films used in this study.

| Number | Chemical formula | Main layer thickness (nm) | Main layer roughness (nm) | Main layer density (g/cm$^3$) | Capping layer | Capping layer thickness (nm) | Capping layer roughness (nm) | $M_s$ (kA/m) | $T_{comp}$ (K) | $H_s$ (kA/m) | $H_c$ (mT) |
|---|---|---|---|---|---|---|---|---|---|---|---|
| 1 | $Dy_{0.25}Co_{0.75}$ | 10.3 | 1.2 | 9.2 | $SiO_2$ | 3.3 | 0.9 | 173.0 | 180 | - | 48.2 |
| 2 | $Dy_{0.25}Co_{0.75}$ | 8.8 | 1.0 | 9.0 | $Al_2O_3$ | 3.2 | 1.0 | 148.5 | 175 | - | 73.0 |
| 3 | $Dy_{0.25}Co_{0.75}$ | 10.5 | 1.6 | 8.2 | $SiO_2$ | 4.6 | 0.8 | 223.0 | 200 | - | 33.6 |
| 4 | $Tb_{0.25}Co_{0.75}$ | 17.1 | 1.6 | 8.0 | $SiO_2$ | 4.5 | 1.5 | 106.0 ; 26.8 * | 200 | - | 295.9 |
| 5 | $Tb_{0.25}Co_{0.75}$ | 20.0 | 1.7 | 7.2 | $SiO_2$ | 4.6 | 0.8 | 120.0 ; 50.7 * | 180 | - | 253.6 |
| 6 | $Tb_{0.20}Co_{0.80}$ | 19.7 | 0.6 | 8.6 | $Al_2O_3$ | 1.5 | 0.5 | 375.0 | 20 | - | 34.0 |
| 7 | $Y_{0.25}Co_{0.75}$ | 8.8 | 1.0 | 6.5 | $SiO_2$ | 2.0 | 0.8 | 705.4 | | 647.0 | < 1.0 |
| 8 | $Y_{0.25}Co_{0.75}$ | 19.0 | 1.3 | 6.5 | $SiO_2$ | 2.0 | 1.5 | 721.7 | | 598.4 | < 1.0 |

*Anhysteretic soft component

## 3. Results.

Many of the Tb films and some of the Dy films showed a soft, anhysteretic component in their magnetization curves that we discuss in § 4.1. We focus first on samples exhibiting a simple square loop and perpendicular magnetic anisotropy.

3.1. Dysprosium alloys

Amorphous a-$Dy_{0.25}Co_{0.75}$ alloys were prepared with a range of thickness from 5 – 94 nm and magnetization was measured both parallel and perpendicular to the film plane. Perpendicular anisotropy was found only in a narrow range of thickness around 10 nm. The alloys exhibit magnetic compensation near 175 K, where the moment of the cobalt subnetwork, measured in a-$YCo_3$ is 4.9 $\mu_B$ [27]. The cobalt moment is practically constant from 4 K to 300 K. A plot in Fig. 3 of the coercivity measured both from magnetization and anomalous Hall loops as a function of temperature shows the divergence at $T_{comp}$. This is confirmed by reversal of the sign of the magnetization of the cobalt subnetwork. The anomalous Hall effect, like the Kerr effect in amorphous rare-earth cobalt alloys, arises essentially from the cobalt magnetization rather than the rare earth. The main evidence for this is that the magnitude of the effect depends little on the alloyed rare earth atom, or on temperature [32]. A recent report on the anomalous Hall effect in the crystalline ferrimagnetic $GdCo_3$ and $GdCo_5$ intermetallics suggests that rare earth and cobalt contributions of opposite sign reinforce each other in the crystalline compounds [34].

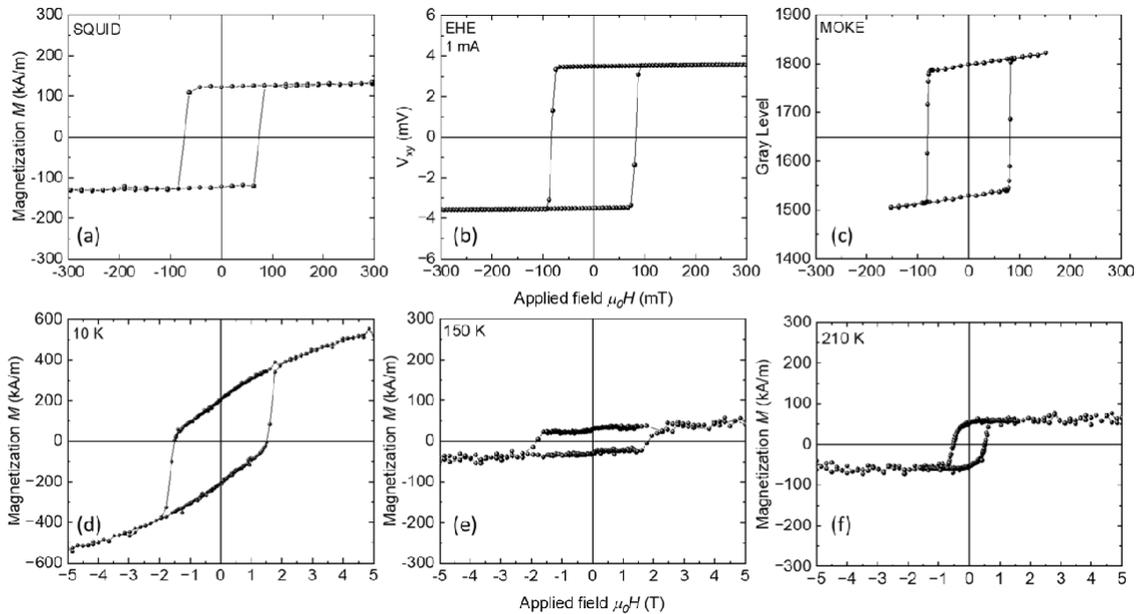

Figure 2 Some representative hysteresis loops of a-$DyCo_3$. Top row: Measured at 300 K a) by SQUID magnetometry perpendicular to the film plane, b) by anomalous Hall effect (AHE) and c) by magneto-optic Kerr effect (MOKE). Bottom row; measured by SQUID magnetometry at d) 10 K, e) 150 K and f) 210 K.

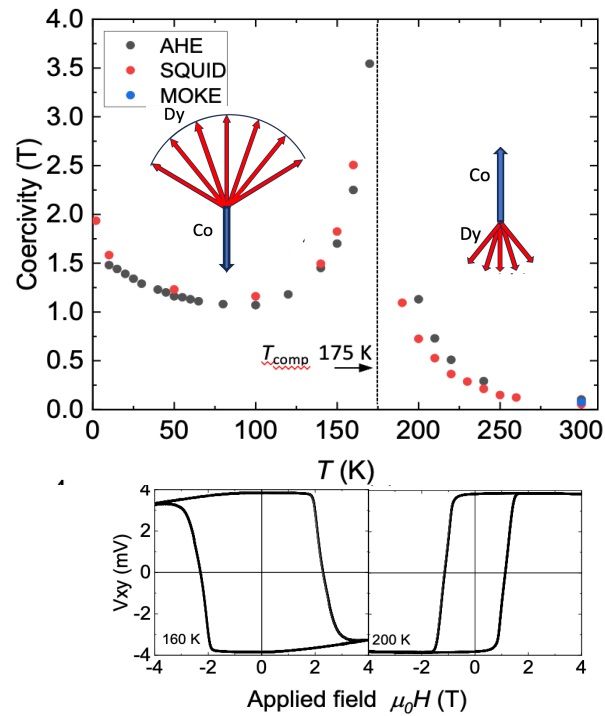

Figure 3 Coercivity for a-$Dy_{0.25}Co_{0.75}$ showing divergence at the compensation temperature of 175 K. Data points are obtained by SQUID magnetometry or AHE. A MOKE point at 300 K is included. Two AHE loops below and above compensation are illustrated in the lower panels.

The remanent magnetic moment $m$ in Bohr magnetons per $DyCo_3$ formula unit is plotted in Fig.4, taking account of the sign change at compensation. The axis on the right shows the net moment per dysprosium atom, assuming the contribution of the cobalt subnetwork remains constant at 4.9 $\mu_B$, and fixing the net Dy moment at this value at $T_{comp}$. There is an upturn in magnetization below 50 K, which is associated with a linear high-field which extrapolates to a magnetization greater than the remanence, as seen in Fig. 2d). At 1.5 K, extrapolation of the high-field magnetization gives a net Dy moment of 7.3 $\mu_B$. The remanence and saturation curves begin to diverge when $K_1$, the net anisotropy of the film, is no longer sufficient to maintain uniform magnetization perpendicular to the film plane, possibly accompanying the onset of surface maze domains. The bulk anisotropy energy of the thin film was estimated as 20 kJm$^{-3}$ at 50 K from the maximum magnetization where the hysteresis loop remains square (see also §4.2) and a similar value, 28 kJm$^{-3}$, was deduced from the 375 mT saturation field $M_s$ in the hard-axis magnetization curve at 300 K using the relation $K_1 = ½ B_a M_s$. The 4f atomic moment per dysprosium atom in the alloy is 10 $\mu_B$, (J = 15/2 and g = 4/3) plus an estimated contribution of 0.5$\mu_B$ from the 5d/6s electrons

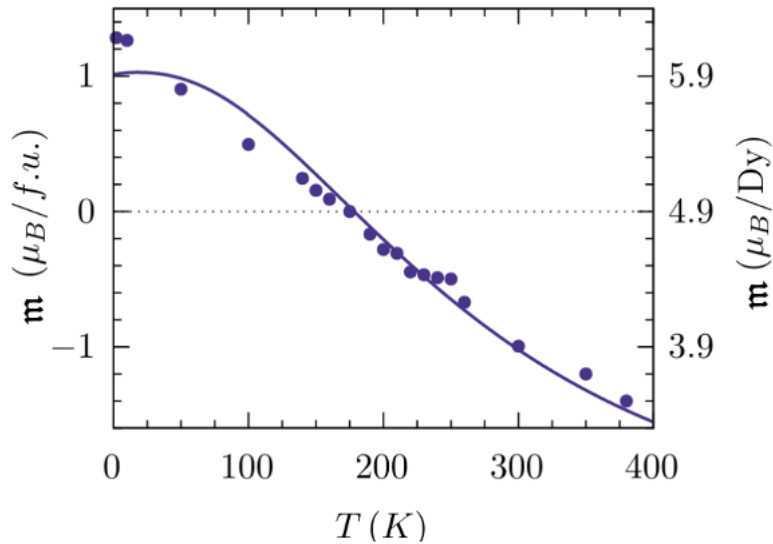

.

Figure 4. Remanent moment per $DyCo_3$ formula unit in a-$Dy_{0.25}Co_{0.75}$, taking account of the change of sign at $T_{comp}$. The magnetic moment per Dysprosium atom is marked on the right hand scale.

A few a-$Dy_xCo_{1-x}$ films exhibit hysteresis loops with two components having coercivities at 300 K of 80 mT and 0 mT, similar to the one illustrated for a-$Tb_{0.25}Co_{0.75}$ in Fig 6d). This is discussed in § 4.1.

A 10 nm DyCo$_3$ film that exhibited partial single-pulse all-optical switching [11] was selected for an X-ray photoemission electron microscopy (XPEEM) investigation on the UE49 PMGa SPEEM beamline at the BESSY II Synchrotron facility at Helmholtz-Zentrum Berlin. The circularly-polarized X-rays were incident at 16° to the sample surface, and the 100 fs laser pulses struck the sample at 10°, producing an elliptical irradiation spot. Images were recorded at the Co L$_3$ (778 eV) absorption edge and x-ray magnetic circular dichroism contrast was calculated from the difference of images taken with left- and right-circularly polarized X-rays, normalized by their sum. Further experimental details are found in reference [35]. The difference maps between subsequent pulses up to the 6th one were recorded and analysed. After the 5th pulse, further net switching was barely detectable However, clear differences were seen in the first few pulse pairs, some of which are illustrated in Figure 5.

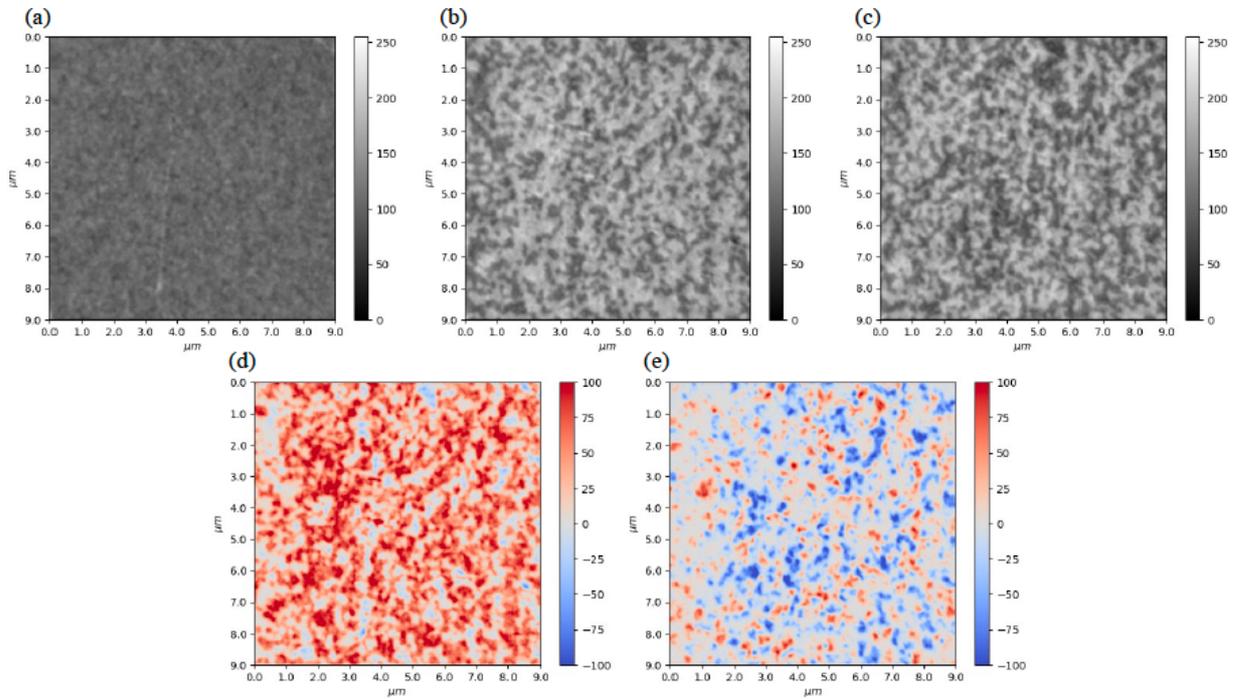

Figure 5 XMCD images of a-Dy$_{0.25}$Co$_{0.75}$ thin film surface at 300 K (a) after saturation in a perpendicular magnetic field, (b) after a first and (c) after a second 100 fs laser pulse. To illustrate the effect of pulsed laser irradiation (d) shows the difference image of (b) and (a) and (e) is the difference image (c) and (b).

3.2 Terbium alloys.

a-Tb$_{0.25}$Co$_{0.75}$ films studied were about 20 nm thick and exhibited perpendicular anisotropy. Most of them show a sharp increase of magnetization in their hysteresis loops close to remanence. An example is shown in Fig. 6d). Remarkably, there is never any trace of this soft phase in either Hall or Kerr hysteresis loops, which are shown for the same film at 300 K in Fig. 6e) and 6f).

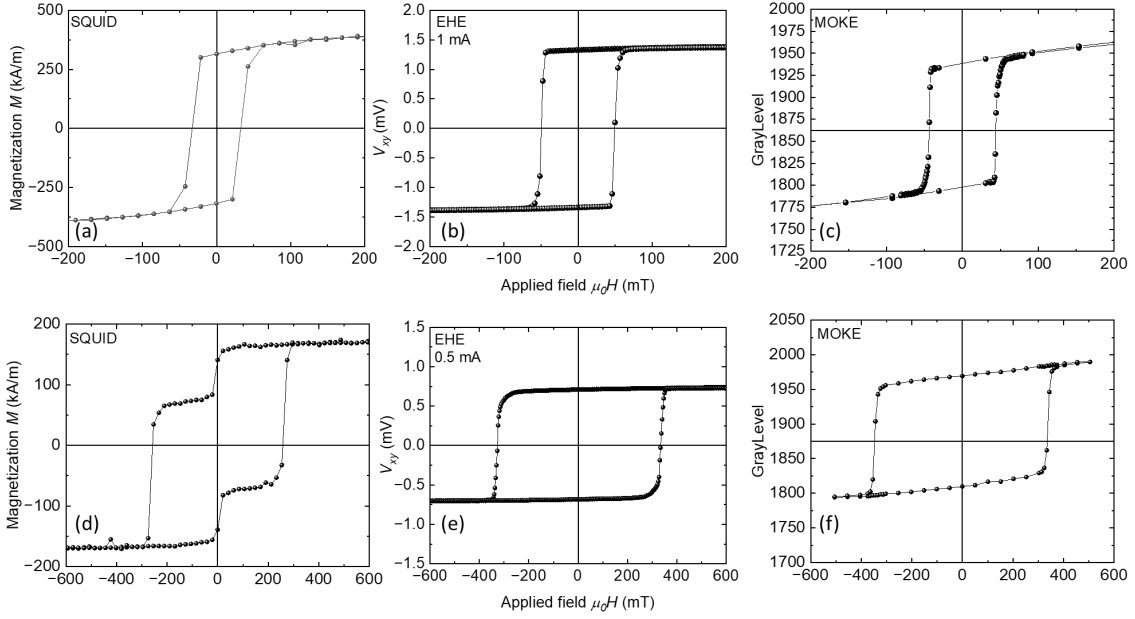

Figure 6. Magnetisation loops measured by SQUID magnetometry (left), anomalous Hall effect (center) and Kerr effect (right) of the a sample of a-$Tb_{0.21}Co_{0.79}$ (top row, film 6) and a-$Tb_{0.25}Co_{0.75}$ (bottom row, film 5). There is never any sign of a step at remanence in the Hall or Kerr loops. All data are at room temperature.

A plot of the coercivity of a-$Tb_{0.25}Co_{0.75}$ as a function of temperature is very similar to that of a-$Dy_{0.25}Co_{0.75}$ in Fig. 3. It tends to diverge at the compensation temperature of about 200 K, where a thermal scan of remanence crosses zero. There is no pronounced appearance of a component with a slope at remanence at lot temperature, possibly because the smaller atomic moment of Tb (9 $\mu_B$) compared to Dy (10 $\mu_B$) limits the magnetization to a value less than the anisotropy field. The net low-temperature remanence of the Tb atoms is 6.8 $\mu_B$ (Fig. 7) and the extrapolated saturation magnetization is 7.0 $\mu_B$. The Terbium film with the larger, temperature-dependent soft component that accounts for about half of the saturation magnetization at room temperature, Fig 6d), exhibited a lower saturation magnetization and compensation near 165 K.

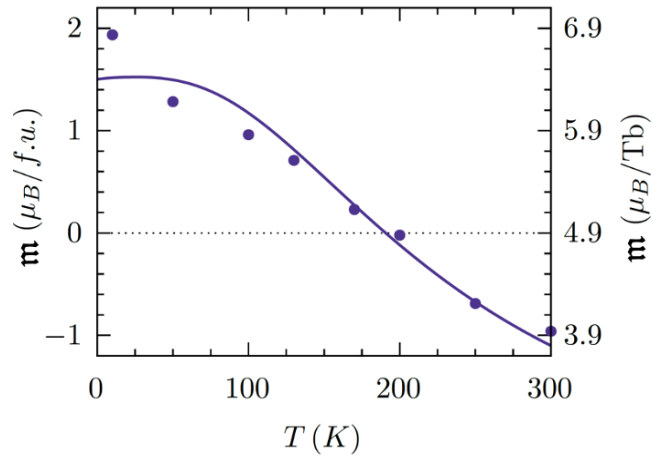

Figure 7. Remanent moment per $TbCo_3$ formula unit in a-$Tb_{0.25}Co_{0.75}$, taking account of the change of sign at $T_{comp}$. The magnetic moment per Tb is shown on the right-hand scale.

## 4. Discussion.

The product of the quadrupole moment of the 4f shell and its interaction with the local electric field gradient, which depends on $J^2$, is just 5% greater for Tb (J = 6) than for Dy (J = 15/2), so the leading second order term in the local random anisotropy is similar for the two elements [30]. The fourth-order term, which may influence the behavior at low temperature, is of opposite sign. The magnetic moments of the 4f shell are also similar for the two atoms, 9 $\mu_B$ and 10 $\mu_B$, respectively [30]. Gd has no 4f quadrupole moment, so it is insensitive to random local electrostatic fields. Its 4f magnetic moment of 7 $\mu_B$ has purely spin character. The amorphous alloys a-$R_x Co_{1-x}$ with R = Gd, Tb, Dy all exhibit compensation near or below room temperature when $x \approx 0.24$. Their spin angular momenta are respectively 3.5, 3 and 2.5 $\hbar$, so the R-Co exchange decreases in this order. The Gd alloys are ferrimagnets, the Tb and Dy alloys are sperimagnets and the a-YCo alloys are ferromagnets since Y is nonmagnetic (Fig. 1) [27].

4.1 The soft component

The soft component of magnetization we observe in a few of the Dy films and most of the Tb films has been reported previously in sputtered a-Tb-Co by several groups [36-40]. It was not observed in the evaporated films that were originally investigated for magneto-optic recording, nor in the early thick sputtered films. Various explanations for the anhysteretic soft magnetization have been advanced. It is not an intrinsic property of the amorphous alloy, because we find it only in some of the films. There are reports of similar effects in as-sputttered rare-earth transition metal multilayers [40]. A soft perpemdicular magnetic layer might make no contribution to the anomalous Hall loops if the corresponding cobalt was electrically insulating, but it should show up in Kerr measurements. It saturates readily (Fig 6d), yet there is no trace of it in the Kerr loops.

A satisfactory explanation of the silent spins that is potentially consistent with all the facts was given by Mandru *et al* [41], who showed that there was no soft component of magnetization in a-$Tb_x Co_{1-x}$ sputtered films provided the 4×4 mm$^2$ chips are freshly cleaved before SQUID measurements, or when a mask is used to prevent deposition of the alloy on the sides of the silicon substrate. Any material there will contribute to the magnetization but will be undetectable in Hall or Kerr measurements.

In our case of 5 × 5 mm$^2$ SQUID samples on 500 µm thick silicon substrates, the area of the four sides is 40 % of the flat area of the chip. Confirmation that a side deposit, and not the thin film itself is the source of the anhysteretic soft magnetization was sought by mechanically polishing the sides of a 5 ×5 mm$^2$ a-$Dy_x Co_{1-x}$ SQUID sample obtained by cleaving a 10 x 10 mm$^2$ sputtered substrate into four quarters. A deposit was visible on two adjacent sides, but not on the sides that had been cleaved. Figure 8 shows that polishing the sides removed the soft component while the remanence of 160 kAm$^{-1}$ was unaltered.

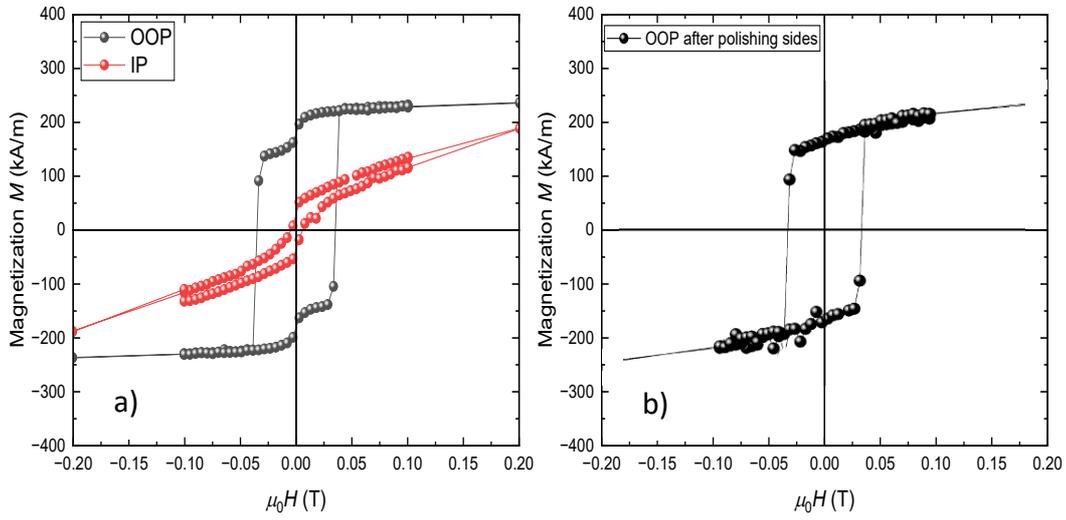

Figure 8. a) Magnetization curves of a 5 × 5 mm$^2$ a-DyCo$_3$ sample (film 3) with an anhysteretic soft component representing 18 % of the magnetization. Magnetization curves are measured in-plane (red) or perpendicular to the plane (black). b) Curve after polishing off the deposit on two adjacent sides.

4.2 Perpendicular magnetic anisotropy.

The magnitude of the intrinsic perpendicular anisotropy in the a-DyCo$_3$ and a-TbCo$_3$ films ($\gtrsim$ 20 kJm$^{-3}$, including the room temperature estimate of 28 kJm$^{-3}$ from the linear in-plane magnetization of film 2 in Fig. 2a)) is smaller than that inferred from the perpendicular magnetization curves of a-Y$_x$Co$_{1-x}$ (46 kJm$^{-3}$) where the large net magnetization lies in-plane [27], which suggests that cobalt rather than the rare earth may be the principal source of perpendicular anisotropy in the Dy and Tb films. Several possible origins of the anisotropy have been discussed in the past, and it is difficult to disentangle them in any particular case. They include surface contributions from the upper and lower surfaces of the thin film, a bulk term associated with substrate-induced strain during deposition, and a structural term due to alignment of R-T pairs at the surface [42], R-Co pairs perpendicular to the film surface [43] or oxygen in the structure [44]. Furthermore, the local easy axes of rare earth atoms at the surface or interface will no longer be random and isotropic. Nanoscale inhomogeneity in the form of columnar rare earth clusters has been demonstrated for a-Gd$_x$Fe$_{1-x}$, and the degree of atomic order or clustering can vary with film thickness [45]. A straightforward empirical separation of bulk and surface contributions based on their scaling with film volume and area is obtainable from a plot of $tK_{\text{eff}}$ vs $t$, where $K_{\text{eff}}$ is the total effective anisotropy and $t$ is the film thickness. This gives the bulk and surface contributions from the slope and y-axis intercept respectively. Applied to a-Tb$_{0.12}$Co$_{0.88}$, the method showed that the surface contribution favours an easy plane and the bulk term an easy axis, with a crossover below 10 nm [32], similar to our a-Dy$_{0.25}$Co$_{0.75}$. However, the bulk anisotropy changes sign in response to some change of composition [46] or amorphous atomic structure with film thickness, which is best approached by empirical exploration of the effect of growth parameters, such as was carried for a-GdFeCo by Hansen and co-workers [7].

## 4.3 Sperimagnetism

The amorphous Tb and Dy alloys are not collinear ferrimagnets, at least in their ground state. If they were, the measured moments at low temperature for a-RCo$_3$ would be 4.1$\mu_B$ and 5.1$\mu_B$ respectively, assuming the collinear cobalt moment is 4.9 $\mu_B$, as measured in a-YCo$_3$ [27] and ignoring any contribution to the rare earth moment from the conduction electrons. The values of about 2.0 $\mu_B$ at low temperature obtained by high-field extrapolation Our observed Dy moment of 7.3 $\mu_B$ is consistent with literature values that range from 6.7$\mu_B$ to 8.5$\mu_B$, depending on the choice of moment for Co and the precise composition and preparation procedure [15, 17-19, 21]. The remanence values (Fig. 4 and Fig. 7) are notably smaller.

The cobalt subnetwork is strongly ferromagnetic and insensitive to local atomic-scale random anisotropy at the cobalt sites of 5.6 K/Co atom because of exchange averaging [27]. The Curie temperature of a-YCo$_3$ is $T_C$ = 760 K [28, 29] and the corresponding molecular field coefficient is $n_W = T_C/C$, where $C = \mu_0 n g^2 \mu_B^2 S(S + 1)/3k_B$ is the Curie constant. The molecular field $H_W = n_W M$, where the magnetization $M = ng\mu_B S$ and n is the number of Co atoms per m$^3$ [30]. It follows that the molecular field depends only on $T_C$ and the effective spin quantum number and g-factor of cobalt.

$$\mu_0 H_W^{Co-Co} = 3k_B T_C/g(S + 1)\mu_B \quad (1)$$

Assuming S = 1 and g = 2, the molecular field $B_W = \mu_0 H_W$ acting on the ferromagnetic cobalt is 855 T, The exchange energy is 1137 K/Co atom, and the ratio α of anisotropy to exchange energy for cobalt is 0.005 [27].

The reduction in the average -z-component rare earth moment of Tb and Dy is attributed to a much greater ratio of random anisotropy energy to exchange anisotropy. In amorphous a-GdCo, where random anisotropy is negligible, the low temperature moment of a-GdCo$_{3.5}$ of 2.0 $\mu_B$ per formula [47] corresponds to the full metallic moment of 7.6 $\mu_B$ per Gd, taking the cobalt moment as 1.6 $\mu_B$. In order to analyse the non-collinearity of Dy and Tb, we need to know the magnitude of the random anisotropy and also the magnitude of $B_W^{R-Co}$, the R-Co exchange field acting on the rare earth, directed opposite to the cobalt magnetization.

A rare earth atom in a magnetically-ordered a-R$_x$Co$_{1-x}$ alloy with uniaxial local anisotropy that does not depend on azimuthal angle is not a small perturbation on the exchange; it is properly described by the one-atom Hamiltonian

$$\mathcal{H}_i = 3B_2^0[(\hat{J}_z \cos\theta_i + \hat{J}_x \sin\theta_i)^2 - (1/3)\hat{J}^2] - g\mu_B \hat{J}_z B_W^{R-Co} \quad (2)$$

where $B_W^{R-Co}$ is the inter-subnetwork molecular field from the cobalt acting on the rare earth. The first term in (2) is the axial second-order 'crystal-field' energy and the second term is the Zeeman energy in the molecular field approximation where g = 2S/J, is the spin g-factor for the heavy rare earth. Here $\theta_i$ is the angle between the magnetization axis of the cobalt subnetwork and the local uniaxial symmetry axis **z´** which is different at every rare earth site. We use the term 'crystal field' to designate the even derivatives of the electrostatic potential created by atomic surroundings of a rare-earth atom that interact with the electronic quadrupole, hexadecapole and 64-pole moments of its 4*f* electron distribution — the second-, fourth- and sixth-order terms. The vector model of Fig. 9 is the classical representation of Eq. 2.

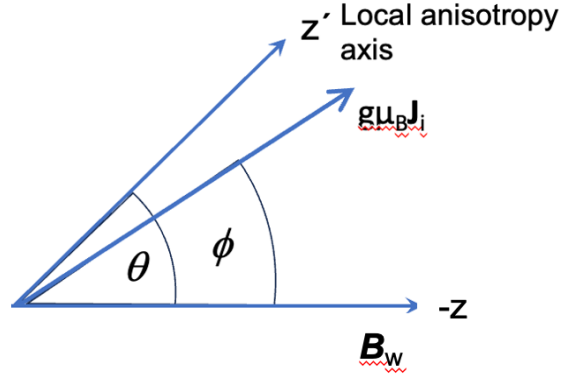

Figure 9. Classical picture of the magnetic moment of a rare earth atom at a site i in an amorphous rare-earth cobalt alloy subject to a molecular field $B_w$ from the cobalt along the -z-axis and uniaxial anisotropy directed along a local axis $z´_i$ making an angle $\theta_i$ with -**z**.

A complete description of the second-order crystal field includes a term describing the asymmetry of the energy in the plane perpendicular to the local easy axis represented by another second-order parameter $B_2^2$ and another, off-diagonal energy term $B_2^2 \hat{O}_2^2$ in the Hamiltonian, which now depends on two angles and involves all three operators $\hat{J}_x$, $\hat{J}_y$, and $\hat{J}_z$ [48]. The local second-order energy surface is a general ellipsoid with three principal axes instead of an ellipsoid of revolution with two, which we consider in Eq.2, neglecting the off-diagonal term $B_2^2 \hat{O}_2^2$.

An approximation valid in the limit when the ratio of anisotropy to exchange energies $\alpha = 3B_2^0 J/g\mu_B B_W^{R-Co}$ is $\ll 1$, replaces the first term in Eq.2 by $B_2^0 \hat{O}_2^0$ where $\hat{O}_2^0 = (3\hat{J}_{zi}^2 - \hat{J}^2)$ is the Stevens operator equivalent for the diagonal term in the electrostatic quadrupole interaction referred to the local easy axis. The small-$\alpha$ approximation is equivalent to representing the leading term in the bulk magneto-crystalline anisotropy by $K_1 \sin^2(\theta - \phi)$. The anisotropy energy is often written as $D\hat{J}_{zi}^2$ with $D = 3B_2^0$, as in the Harris, Plischke, Zuckerman Model model [49], but the $\hat{J}^2$ term with diagonal elements J(J+1) is needed to ensure that anisotropy vanishes at high temperatures when all the $M_j$ states of a single ion are equally populated. The local anisotropy when $\alpha \ll 1$ is exchange-averaged by the molecular field, just as it is in a-$Y_xCo_{1-x}$, but when $\alpha \gtrsim 1$, the local magnetization direction approaches **z´** and $(\theta - \phi)$ becomes small; the small-$\alpha$ approximation is no longer valid because $\sin^2(\theta - \phi)$ is then always $< \cos \theta$.

We assume furthermore that $B_W^{R-Co}$ and $B_2^0$ have constant magnitude at every site and are independent of temperature. The values of $B_W^{R-Co}$ and $B_2^0 = D/3$ are then needed to define α. The anisotropy parameter has been estimated experimentally for rare-earth rich amorphous R-X alloys, where R = Tb or Dy and X = Cu, Ag or Au which exhibit spin freezing at temperatures in the range 20 – 70 K [50]. For example, a value of $D$ = 3 K was deduced from the large linear low-temperature specific heat of a-$Dy_{0.41}Cu_{0.59}$ [51]. Similar estimates of $D$ were based on the approach to saturation of a-$R_{0.50}Ag_{50}$ alloys [52]. The electric field gradient at the rare earth site in transition-metal rich a-$RT_3$ alloys should be greater than in a-RX because of the shorter average nearest-neighbor distances. More realistic estimates may be obtained from a-$RNi_3$ [53], where the Ni itself is nonmagnetic, but similar in size to Co. Random spin freezing as a result of weak R-R exchange and random anisotropy occurs at 10 K for R = Dy. A fit of the linear magnetization curve at 1.4 K in high fields from 8 - 15 T, where the moment per Dy is about 6 $\mu_B$ gave $D$ = 5.8 K [53]. The corresponding spin freezing temperature and $D$ for Tb would be 12 K and 6.1 K, scaling by the rare earth spin and quadrupole interaction respectively [30]. Substantially-larger values of $D$ were reported in sperimagnets where the rare earth is simultaneously subject to random anisotropy and a large molecular field $B_W^{R-Co}$ from cobalt [48, 54, 55].

In Figure 10 we show a plot of reduced magnetization $\langle J_z \rangle/J$ versus α over the range 0 < α < 4, calculated from Eq 2, where $\langle J_z \rangle/J = \langle m \rangle/m_o$ is evaluated for different values of α by diagonalizing the one-atom Hamiltonian of Eq. 2 and averaging over the axes in a hemisphere. The plot is for J = 15/2 (Dy) but the curve for J = 6 (Tb) is practically the same.

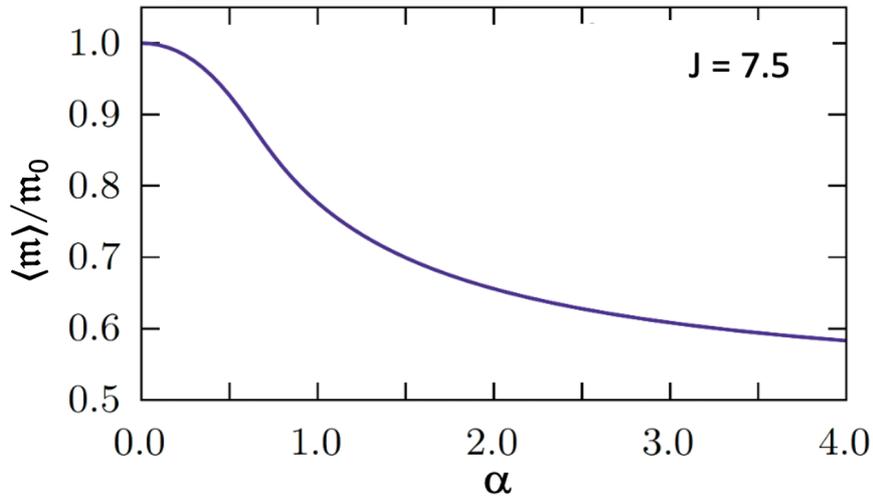

Figure 10. Reduced-low temperature magnetization $\langle m \rangle/m_0$ calculated from Eq 2 for a Dy or Tb atom with random anisotropy as a function of α, the ratio of anisotropy to exchange energy per atom.

When α ≪ 1, we recover the classic approach to saturation for randomly-oriented uniaxial particles varying approximately as $1/B^2$, which was recently re-evaluated by Hui Zhang *et al* [56]. To compare the measured low-temperature values of $\langle J_z \rangle/J$ of around 0.7 for Dy and Tb with the theoretical predictions, we need α values for a-DyCo$_3$ and a-TbCo$_3$. The exchange energy can be deduced from the compensation in a-GdCo$_3$, which is at 325 K [47]. The exchange field required to induce a Gd magnetization of 4.9 μ$_B$, the Co moment at this temperature [27], is 140 T for a Brillouin function $\mathcal{B}_{7/2}(x)$ with g = 2. Scaling by the ratio of the spin moments of Dy/Gd (5/7), the exchange field acting on the Dy subnetwork is 100 T. The random anisotropy prevents the exchange field from aligning the Dy moments along -**z** so the corresponding magnetic energy is less than that expected for the moment calculated from the Brillouin function. A fit of the theory of Eq. 2 to the temperature dependence of the remanence of the Dy alloy in Fig. 4, assuming a moment distribution P{$m(T,\theta)$}, and fixing $T_{comp}$ = 175 K, gives the temperature-dependence shown in Fig. 4. A similar analysis for Tb gives the fit shown in Fig. 7. The corresponding values of $D$ and $B_W^{R-Co}$ are 19 K and 90 T for Dy and 12 K and 77 T for Tb. The values of α and the cone angles at $T = 0$ are 3.6 and $\theta_0$ = 81° for Dy and 1.4 and 67° for Tb. However the calculated temperature-dependence of the rare earth moment, which has zero slope at $T = 0$, deviates obviously from the experimental remanence data below about 50 K. The model gives moments $\langle m(0) \rangle$ of 5.9 μ$_B$ for Dy and 6.4 μ$_B$ for Tb, which are 0.3 or 0.5 μ$_B$ lower that the remanence extrapolated to $T = 0$ and 1.4 or 0.9 μ$_B$ less than the high-field extrapolations.

The model of Eq.2, based on a uniaxially symmetric second-order local crystal field parameterised by a single constant $B_2^0$ or $D$ is undoubtedly oversimplified. There is another second-order parameter $B_2^2$ that reflects the asymmetry of the crystal field energy in the local x´- y´ plane, as well as fourth- and sixth-order terms that reflect crystal field components with cubic or hexagonal symmetry. The higher-order diagonal crystal field terms with opposite sign to the second-order term (fourth-order for Tb and sixth order for Dy) make the local energy surface more spherical and reduce the energy for thermal excitation. The importance of the higher-order terms at low temperature is recognized in crystalline rare earth metals and intermetallics such as Nd$_2$Fe$_{14}$B [3], which becomes noncollinear with four-fold magnetic symmetry below 135 K. Unlike a-YCo, where the random anisotropy is exchange-averaged because α ≪ 1, both a-DyCo$_3$ and a-TbCo$_3$ are in a strong pinning region, α ≥ 1 where the local anisotropy is not a small perturbation, but dominates the exchange. The coercivity of several tesla at low temperature in these amorphous magnets is a result.

The effect of random uniaxial anisotropy $DJ^2$ of order several hundred K per rare earth atom is to reduce the accessible magnetic states at low temperature to just two, ±M$_{Jz'}$ along the local easy axis z´. They become random Ising moments, rather than random Heisenberg moments, and are much easier to order magnetically. Any exchange interaction with the cobalt subnetwork at low temperature, however small, will tend to create a uniform distribution of moments within a hemisphere, which narrows to a cone in increasing exchange field. Furthermore, paramagnetic fluctuations are inhibited at low temperature by the large change of angular momentum required for a spin flip from $M_J$ to $-M_J$ states. They are magnetic 'low entropy alloys' thanks to the action of the local crystal field.

The magnitude of the rare-earth moment is reduced by thermal excitation to excited $M_J$ states on increasing temperature, but the anisotropy is simultaneously weakened, thereby narrowing the cone. For Dy in $DyCo_3$ for example, with $\alpha = 3.6$, the local atomic moment and cone angle at 400 K are reduced to 3.4 $\mu_B$ and 71º, and the sperimagnetic structure progressively approaches collinear ferrimagnetism with increasing temperature. The temperature sequence of magnetic structure as a function of temperature in an a-$RCo_3$ film with perpendicular magnetic anisotropy is illustrated schematically in Fig. 12 below.

4.4 Compensation.

The behaviour of a ferrimagnet or sperimagnet with perpendicular anisotropy at compensation is illustrated by data on different rare earths. Simplest are ferrrimagnetic a-$Gd_xCo_{1-x}$ and a-$Gd_x(Fe,Co)_{1-x}$ [7]. Although the films exhibit perpendicular anisotropy with a dominant Gd subnetwork below $T_{comp}$ and a dominant Co subnetwork above, the perpendicular magnetic configuration becomes unstable in an out-of-plane magnetic field when the two subnetwork magnetizations compensate exactly [34, 58, 59], The situation is similar to an antiferromagnet in a field applied parallel to the antiferromagnetic axis (Néel vector). The antiferromagnet undergoes a spin flop transition to a canted configuration in a critical field $H_{sf}$ when the Zeeman energy in a perpendicular field is equal to the anisotropy energy; $H_{sf}$ is given by the expression [30]

$$H_{sf} = 2\sqrt{(H_a H_{ex})} \qquad (4)$$

when $T_{sf}$ is much less than the Néel temperature $T_N$ so that the parallel susceptibility can be neglected. The fields $H_a$ and $H_{ex}$ are the anisotropy field and the inter-subnetwork exchange field. Bigger applied fields can flop the magnetization of a ferrimagnet in a range of temperature on either side of compensation, provided the net magnetization is small, as it is in a-GdCo [57,59], crystalline ferrimagnets [60] and amorphous sperimagnets near compensation. Ferromagnetic saturation would require a much larger applied field, of order $H_{ex}$. If $M_\alpha$ is the magnetization of either subnetwork, the perpendicular susceptibility of the antiferromagnet is

$$\chi_\perp = M_\alpha/H_{ex}$$

so the net magnetization in an applied field $H$ is $M_\alpha H/H_{ex}$. The parallel susceptibility can be neglected when $T_{comp} \ll T_c$

Figure 11 shows measurements of the anomalous Hall effect of sperimagnetic a-DyCo$_3$ in fields of 14 T in the temperature range from 10 – 300 K, together with four hysteresis loops obtained at temperatures below or above compensation. Unlike an antiferromagnet, where there is a single, well-defined spin flop in a field $H_{sf}$ that increases with decreasing temperature, a ferrimagnet or sperimagnet shows a spin flop at $T_{comp}$ in a field $H_{sf}^0$ and extended spin flops with an onset that increases linearly with $|T - T_{comp}|$ and begins at $H_{sf}^0$. The onset of the extended spin flop of the cobalt subnetwork is indicated by an arrow in the 200 K hysteresis loop. The plot in the insert to Fig. 11 of the onset spin flop field $\mu_0 H_{sf}$ versus temperature with slope 0.15 T/K extrapolates to $\mu_0 H_{sf}^0 = 2.0$ T at $T_{comp} = 175$ K. The slope $M$ of versus $| H_{sf} - H_{sf}^0|$ is 9 ×10$^{-3}$. We can estimate $H_{sf}^0$ from (4) if we know $H_{ex}$ and $H_a$. The inter-subnetwork exchange field $\mu_0 H_{ex}$ was shown in § 4.3 to be 90 T. Taking $K_1 = 20$ kJm$^{-3}$ and $M_\alpha = 600$ kAm$^{-1}$, we have $\mu_0 H_a = 0.0167$ T. The value of $\mu_0 H_{sf}^0$ estimated from Eq 4 is 2.4 T, in acceptable agreement with the measured extrapolation. The spin flop is abrupt at compensation as it is in an antiferromagnet, but when there is a net magnetization — in a-DyCo$_3$ the net magnetization $M$ near compensation increases at a rate of 17 kAm$^{-1}$/K — the Co flop nucleates in a field greater than $\mu_0 H_{sf}^0$, and more than 15 T is needed to complete the flop at 200 K where $M$ is 250 kAm$^{-1}$.

A feature of the data in Fig. 11 is that the deviation from saturation of the cobalt magnetization measured by AHE sets in close to zero field below compensation, and above it above compensation. This is unlike a-Gd(Fe,Co) where the deviation from saturation is symmetric below and above $T_{comp}$ [57, 58]. The asymmetry reflects the dominant magnetic subnetwork, Dy below and Co above compensation. Whereas the magnitude of the Co subnetwork moment is insensitive to applied field in this temperature range, the cone angle of the Dy will be reduced with increasing temperature; the effect of an applied field of 5 T will be to increase or decrease the Dy average moment by about 5 %, respectively. This may be the reason for the asymmetry.

A recent study of the anomalous Hall effect at compensation in crystalline GdCo$_5$ and GdCo$_3$ intermetallic compounds [34] questioned the received view that the rare earth contribution to the AHE is negligible. The sign of the AHE in elemental Gd is opposite to that in metallic Co, and it is therefore possible that the 5d electrons of Gd, Tb or Dy contribute in the same sense as Co to the AHE. We do not think this is a significant effect in our amorphous alloys, for two reasons. Firstly there is little change in the magnitude of the AHE between 10 K and 300 K. It decreases by 3 %, whereas the rare earth moment falls by about 50 % in the same range. Secondly, the magnitude of the Hall voltage is similar in a-TbCo$_3$ and a-YCo$_3$.

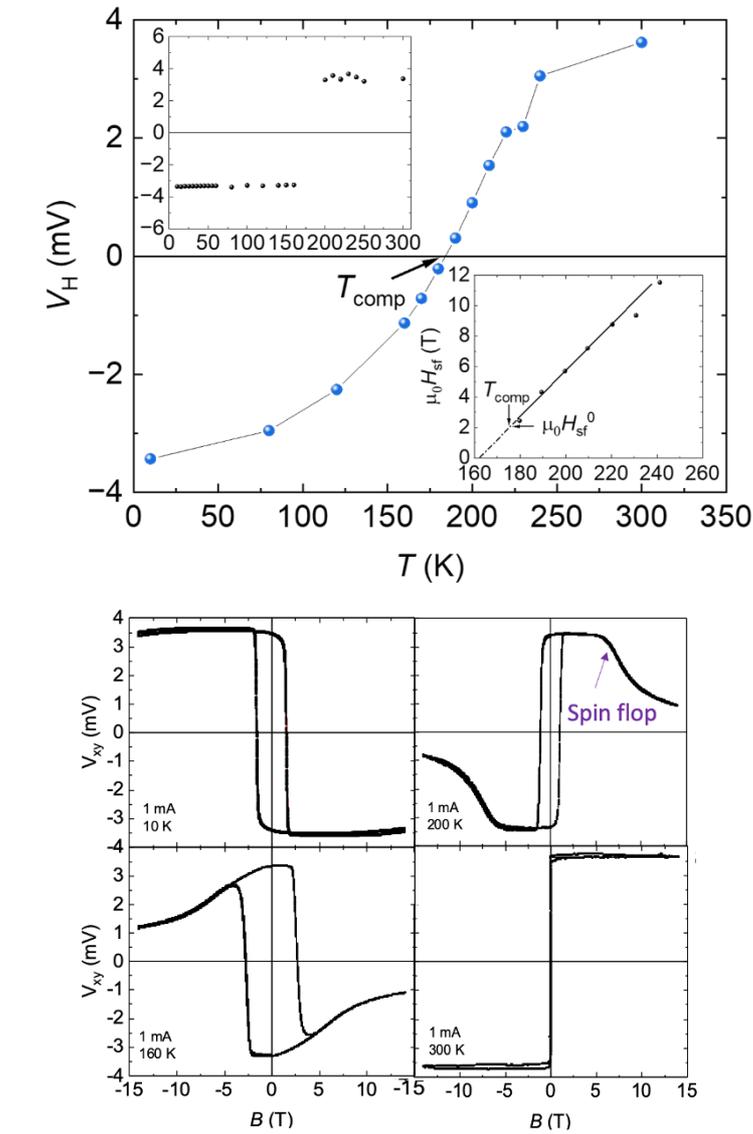

Figure 11. Anomalous Hall voltage for a-DyCo$_3$ measured in 14 T with a current of 1 mA; examples of complete Hall loops are shown in the four panels. The decreases in high fields above and below $T_{comp}$ = 175 K mark the onsets of extended spin flops. Inserts show the temperature dependence of the onset of the extended spin flop above 175 K, and the temperature dependence of the anomalous Hall effect at remanence.

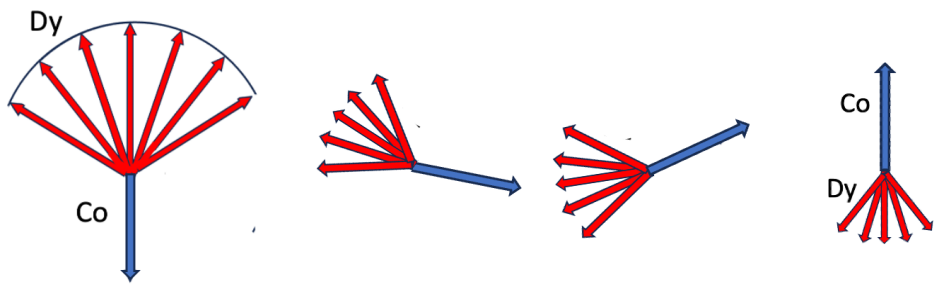

Figure 12. The sequence of magnetic structures found with increasing temperature in a-DyCo$_3$ on passing through compensation in a field greater than the spin-flop field.

## 4.5 Single-pulse all-optical switching.

Our study of partial single-pulse all-optical toggle switching by XPEEM in a-DyCo$_3$ confirms that the effect is observed both above and below $T_{comp}$, but the laser power required to switch the domain structure increases, like the local anisotropy, with decreasing temperature. The ratio of the laser power needed to switch increases by a factor 1.9 from 130 mW to 250 mW on cooling across compensation from 200 K to 120 K, whereas the Dy anisotropy energy increases by a factor of 1.8 from 15.8 K to 29.1 K in the same temperature range. It is the magnitude of the local rare earth anisotropy that controls the switch, not the macroscopic anisotropy field.

The domains depends somewhat on pulse energy at a given temperature, and there is always a variety domain sizes in the range 100 – 800 nm. Domains are not fixed regions of the film that switch, but they appear in different positions every time. The switching appears to be uncorrelated with defects or spatial inhomogeneity in the film.

It is interesting that complete single-pulse all-optical toggle switching is a purely thermal effect that has only been found in ferrimagnetic a-RT alloys that contain Gd [61], or Mn-based crystalline ferrimagnetic metals [62]. The common feature of these systems is an element with a very stable ground-state atomic multiplet corresponding to a half-filled shell. $4f^7$ or $3d^5$; no other multiplet that can be populated by optical excitation. Hund's third rule is respected in the excitation process.

## 5. Conclusion

The rare earth contribution to the low temperature magnetization of a-DyCo$_3$ and a-TbCo$_3$ is reduced at saturation to about 70% of the value expected from the *4f* atomic moment, by a noncollinear magnetic structure attributed principally to random uniaxial magnetic anisotropy at the rare earth sites and a negative sign of the quadrupole moment of the 4f charge distribution. At very low temperature, the Ising-like rare earth moments will be aligned in directions lying within a hemisphere in a weak unidirectional exchange field arising from the cobalt subnetwork; increasing R-Co exchange narrows the hemisphere to a cone, a sperimagnetic structure comprising a noncollinear rare earth subnetwork with an antiparallel ferromagnetic cobalt subnetwork. The ratio α of the average magnitude of the anisotropy to the exchange per atom is 3.4 for Dy ($B_2^0$ = 6.5 K and $B_W^{R\text{-}Co}$ = 90 T) and value for Tb is 1.4 ($B_2^0$ = 4 K and $B_W^{R\text{-}Co}$ = 77 T). The rare-earth cone angles at T = 0 are 81º and 67º, respectively, and they are reduced by thermal excitation as the temperature increases but the magnetic structure approaches collinear ferrimagnetism only at temperatures well above 300 K.

The modelling of the random magnetic structures assuming the magnitude of $B_2^0$ is the same at every site and neglecting any R-R exchange and all other terms other terms in the 'crystal field' acting on the rare earth is oversimplified and clearly fails at temperatures below about 50 K. Detailed numerical calculations on large amorphous binary atomic models are needed to better characterize the distribution of the local electrostatic potential and higher-order 'crystal field' contributions to the anisotropy. Tb and Dy have the largest random uniaxial anisotropy; the negative quadrupole moment for Ho is smaller; Er, Tm and Yb experience weaker R-Co exchange due to their smaller spins, and their uniaxial anisotropy has the opposite sign — easy-plane rather than easy-axis leading to quite different magnetic behavior. Calculations of the effects of the electrostatic interactions on a-$R_{1-x}Co_x$ alloys with rare earths having a range of different magnetic dipole and electric multipole moments of both signs are needed to improve the oversimplified model. Other open questions are whether the ground-state magnetic structure evolves continuously from weak to strong pinning as a function of $\alpha$ or whether there is an abrupt transition, as suggested by early small-scale computer simulations [63] and the role of random anisotropy, rather the Dzyaloshinskii-Moria interaction in favouring skyrmions in the weak-pinning limit [24,25]. It seems that higher order crystal field terms may be critical for determining the magnetic order at low temperatures in amorphous rare earth transition-metal alloys, just as they are in the crystalline counterparts.

**Acknowledgements** This work was supported by Science Foundation Ireland under grants 16/IA/4534 ZEMS, 12/RC/2278 AMBER, 17/NSFC/5294 MANIAC and EU FET Open grant 737038 TRANSPIRE. KS acknowledges a postdoctoral fellowship from the Humboldt Foundation. ZH acknowledges support of a Postgraduate Scholarship from Trinity College Dublin. We are grateful to Florian Kronast and Alevtina Smekhova for support during the synchrotron experiment, to Yangkun He for help with the magnetization measurements and to Simon Lenne for help with the Hall effect.

**References**

[1] R.E. Wallace, *Rare Earth Intermetallics*, Elsevier, 2012.
[2] K. Buschow, Intermetallic compounds of rare-earth and 3d transition metals, Reports on Progress in Physics, 40 (1977) 1179.
[3] J.M.D. Coey, *Rare-earth Ion Permanent Magnets*, Oxford University Press, 1996.
[4] P. Hansen, Magnetic amorphous alloys in Handbook of Magnetic Materials, vol 6 (K.H.J. Buschow, Editor) North Holland, Amsterdam, 1991 Ch. 4, 289-452
[5] J.M.D. Coey, Amorphous magnetic order, Journal of Applied Physics, 49 (1978) 1646-1652.
[6] M. Mansuripur, *The Physical Principles of Magneto-optical Recording*, Cambridge University Press 1995
[7] P. Hansen, C. Clausen, G. Much, M. Rosenkranz, K. Witter, Magnetic and magneto-optical properties of rare-earth transition-metal alloys containing Gd, Tb, Fe, Co, Journal of Applied Physics, 66 (1989) 756-767.
[8] C. D. Stanciu, F. Hansteen, A. V. Kimmel, A. Kirilyuk, A. Tsukamoto, A. Itoh and Th Rasing. All-optical magnetic recording with circularly polarized light, Physical Review Letters, 99 (2007) 047601
[9] T. Ostler, J. Barker, R. Evans, R. Chantrell, U. Atxitia, O. Chubykalo-Fesenko, S. El Moussaoui, L. Le Guyader, E. Mengotti, L. Heyderman, Ultrafast heating as a sufficient

stimulus for magnetization reversal in a ferrimagnet, Nature Communications, 3 (2012) 666.

[10] A. El-Ghazaly, B. Tran, A. Ceballos, C.-H. Lambert, A. Pattabi, S. Salahuddin, F. Hellman, J. Bokor, Ultrafast magnetization switching in nanoscale magnetic dots, Applied Physics Letters, 114 (2019) 232407.

[11] Z. Hu, J. Besbas, R. Smith, N. Teichert, G. Atcheson, K. Rode, P. Stamenov, J.M.D. Coey, Single-pulse all-optical partial switching in amorphous $Dy_xCo_{1-x}$ and $Tb_xCo_{1-x}$ with random anisotropy, Applied Physics Letters, 122 (2023) 022401.

[12] Liu, T.-M., Wang, T., Reid, A. H., Savoini, M., Wu, X., Koene, B., Granitzka, P., Graves, C. E., Higley, D. J. & Chen, Z. "Nanoscale confinement of all-optical magnetic switching in TbFeCo-competition with nanoscale heterogeneity". Nano letters **15**, 6862 (2015).

[13] Y. Liu, H. Cheng, P. Vallobra, H. Wang, S. Elmer, X. Zhang, G. Malinowski, M. Hehn, Y. Xu, S. Mangin, W. Zhao, Ultrafast single-pulse switching of Tb-dominant CoTb alloy. Applied Physics Letters, 120 (2022) 022401.

[14] S. Tsunashima, S. Masui, T. Kobayashi and S. Uchiyama, Magneto-optic Kerr effect of amorphous Gd-Fe-Co films. Journal of Applied Physics, 53 (1989) 5175-5177.

[15] S. Uchiyama, Magnetic properties of rare earth – cobalt amorphous films, Materials Chemistry and Physics, 42 (1995) 38-44

[16] P. Hansen, S. Kahn, C. Clausen, G. Much, K. Witter, Magnetic and magneto-optical properties of rare-earth transition-metal alloys containing Dy, Ho, Fe, Co, Journal of Applied Physics, 69 (1991) 3194-3207.

[17] J.M.D. Coey, J. Chappert, J. Rebouillat, T. Wang, Magnetic structure of an amorphous rare-earth transition-metal alloy, Physical Review Letters, 36 (1976) 1061-1064

[18] K. G. Balymov, F. V, Kudyukov, V. O. Vas'kovskiy, O. A. Adankova, N. A. Kulesh, E. A. Stepanova, A. S. Rusalina, Magnetism of amorphous Dy-Tb-Co-type films J. Phys. Conf. Series, 1389 (2019) 012014

[19] R. Hussain, Aakansha, B. Bhrama, B. K. Basumatary, R. Bhrama, S. Ravi, K. Shrivastava, Sperimagnetism in perpendicularly magnetized Co-Tb alloy-based thin films, J. Superconductivity and Novel Magnetism 32 (2019) 4027 - 4031

[20] V. O. Vas'kovskiy, E. V. Kudyukov, E. A. Stepanova, E. A. Kravtsov, O. A. Adankova, A. S. Rusalina, K. G. Balymov, A. V. Svalov, Physics of MAeals and Metallography 122 (2021) 478-484

[21] K. Chen, D. Lott, F. Radu, F, Choueikani, E. Otero and P.Ohresser, Temperature-dependent magnetic properties of ferrimagnetic $DyCo_3$ alloy films, Physical Review B 92 (2015) 024409

[22] D. H. Suzuiki, M. Valvidares, P. Gargiani, M. Huang, A. E. Kossak and G. S. D. Beach, Thickness and composition effects on atomic moments and magnetic compensation point in rare-earth transition metal thin films, Physical Review B 107 (2023) 134430

[23] D-H Kim, M. Haruta, H-Y Ko, G. Go, H-J Park, T. Nishimura, D-Y Kim, T. Okno, Y. Hirata, H. Yoshikawa, W. Ham, S. Kim, H. Kurata, A. Tsukamoto, Y. Shiota, T. Moriyama, S-B Choe, K-J Lee, T. Ono, Bulk Dzyaloshinskii–Moriya interaction in amorphous ferrimagnetic alloys, Nature Materials 18 (2019) 685–690

[24] K. Chen, D. Lott, A. Philppi-Kobs, M. Weigrand, C. Luo and F. Radu, Observation of compact ferrimagnetic skyrmions in a $DyCo_3$ film, Nanoscale 12 (2020) 18137-18143

[25] C. Luo, K. Chen, V. Ukleev, S. Wintz, M. Weigand, R-M, Abrudan, K. Prokeš, F. Radu, Direct observation of Néel-type skyrmions and domain walls in a ferrimagnetic $DyCo_3$ thin film, Communications Physics 6 (2023) 218.

[26] S. K. Kim, G. S. D. Beach, K-J Lee, T Ono, T. Rasing and H. Yan, Ferrimagnetic spintronics, Nature Materials 21 (2022) 24-34

[27] Z. Hu, J. Besbas, K. Siewierska, R. Smith, P. Stamenov and J. M. D. Coey, Magnetism, transport and atomic structure of amorphous binary $Y_xCo_{1-x}$ alloys. Physical Review B (in press).
[28] Y. Kakehashi, M. Yu, Metallic magnetism in amorphous Co-Y alloys, Physical Review B, 49 (1994) 15076.
[29] K. Fukamichi, T. Goto, U. Mizutani, The crystallization temperature, electrical resistivity and magnetic properties of Co-Y amorphous alloys, IEEE transactions on Magnetics, 23 (1987) 3590-3592.
[30] J.M.D. Coey, Magnetism and Magnetic Materials, Cambridge University Press, 2010
[31] G. Adachi and N. Imanaka, The Binary Rare Earth Oxides, Chemical Reviews 98 (1998) 479-1414
[32] R. Asomoza, I. A. Campbell, H. Jouve, R. Meyer, Extraordinary Hall effect in rare-earth–cobalt amorphous films, Journal of Applied Physics 48 (1977) 3829–3831
[33] D. Hou, Y. Li, D. Wei, D.Tian, L.Wu and X. Jin, The anomalous Hall effect in epitaxial face-centered-cubic cobalt films, Journal of Physics: Condensed Matter 24 (2012) 482001
[34] N. T. Hai, J-C Wu, J-P Chou and J. Pothan, Novel anomalous Hall effect mechanism in ferrimagnetic Gd-Co alloys, Journal of Applied Physics 133 (2023) 233901
[35] A. Arora, M.-A. Mawass, O. Sandig, Chen Luo, A. A. Ünal, F. Radu, S. Valencia, F. Kronast, Spatially resolved investigation of all optical magnetization switching in TbFe alloys, Scientific Reports 7 (2017) 9456
[36] A. Ceballos, M. Charilaou, M. Molina-Ruiz, F. Hellman, Coexistence of soft and hard magnetic phases in single layer amorphous Tb–Co thin films, Journal of Applied Physics, 131 (2022) 033901.
[37] K. A. Thorarinsdóttir, B. R. Thorbjarnarsdóttir, U.B. Arnalds and F. Magnus. Competing interface and bulk anisotropies in Co-rich TbCo amorphous films. Journal of Physics, Condensed Matter 35 (2023) 205802
[38] Ji-Ho Park, Won Tae Kim, Woonjae Won, Jun-Ho Kang, Soogil Lee, Byong-Guk Park, Byoung S. Ham , Younghun Jo, Fabian Rotermund, Kab-Jin Kim, Observation of spin-glass-like characteristics in ferrimagnetic TbCo through energy-level selective approach, Nature Communications 13 (2022) 5530
[39] Ke Wang, Ya Huang, Zhan Xu, Shuo Dong, Ruofei Chen, Effect of sputtering power on the magnetic properties of amorphous TbFeCo films, Journal of Magnetism and Magnetic Materials, 424 (2017) 89-92
[40] Siwei Zhang, Yu Zhang, Linlin Zhang, Ziyang Li, Yang Ren, Qingyuan Jin, Zongzhi Zhang, Temperature Dependence of Magnetic Properties in CoFe/Tb Multilayers with Perpendicular Magnetic Anisotropy, ACS Applied Electronic Materials, 4 (2022) 5361-5367
[41] Andrada-Oana Mandru, Oğuz Yıldırım, Miguel A. Marioni, Hartmut Rohrmann, Michael Heigl, Oana-Tereza Ciubotariu, Marcos Penedo, Xue Zhao, Manfred Albrecht, Hans J. Hug, Pervasive artifacts revealed from magnetometry measurements of rare earth-transition metal thin films. Journal of Vacuum Science and Technology A 38 (2020) 023409.
[42] Hong Fu, M. Mansuripur, P. Meystred Generic Source of Perpendicular Anisotropy in Amorphous Rare-Earth-Transition-Metal Films Physical Review Letters, 66 (1991) 1086
[43] V. G. Harris, K. D. Aylesworth, B. N. Das, W. T. Elam, and N. C. Koon, Structural Origins of Magnetic Anisotropy in Sputtered Amorphous Tb-Fe Films, Physical Review Letters 69 (1992) 1939

[44] K. Srinivasan, Yulan Chen, L. Cestarollo, D. K. Dare, J. G. Wright, A.El-Gha Engineering large perpendicular magnetic anisotropy in amorphous ferrimagnetic gadolinium cobalt alloys, Journal of Materials Chemistry C 11 (2023) 4820-4829

[45] E. Kirk, C. Bull , S. Finizio, H. Sepehri-Amin, S. Wintz , A. K. Suszka, N. S. Bingham, P. Warnicke, K. Hono, P. W. Nutter, J. Raabe , G. Hrkac, T. Thomson, L. J. Heyderman, Anisotropy-induced spin reorientation in chemically modulated amorphous ferrimagnetic films, Physical Review Materials 4 (2020 ) 077403

[46] Ke Wang, Zhan Xu, Shuo Dong, A simple method for tuning perpendicular magnetic properties of ultra-thin TbFeCo films, Materials Letters 236 (2019) 89-91

[47] A. Gangulee and R. C. Taylor, Mean Field analysis of the magnetic properties of vapour deposited amorphous Gd-Fe thin films. Journal of Applied Physics 49 (1978) 1762-1764.

[48] R. W. Cochrane, R. Harris and M. J. Zuckermann, The role of structure on the magnetic properties of amorphous alloys, Physics Reports **48** (1978) 1-63

[49] R. Harris, M. Plischke, M. Zuckermann, New model for amorphous magnetism, Physical Review Letters, 31 (1973) 160.

[50] K.Moorjani and J.M. D. Coey, Magnetic Glasses, Elsevier, Amsterdam 1986 pp 266-298

[51] J. M. D. Coey and S. von Molnar, Linear specific heat of on amorphous magnet due to single-ion excitations, Journal de Physique Lettres 39 (1978) L327-330

[52] B. Boucher, Bulk magnetic properties of the amorphous magnetic alloys $RE_{50}Ag_{50}$ (RE = Gd, Tb, Dy, Ho and Er) IEEE Transactions on Magnetics 13 (1977) 1601-1602

[53] R. Asomoza, I. A. Campbell, A. Fert, A. Liénard and J. P. Rebouillat, Magnetic and transport properties of nickel-rare earth amorphous alloys .J. Physics F, Metal Physics **9** (1979) 349-372

[54] A.K. Bhattacharjee, B. Coqblin, R. Jullien, M.J. Zuckermann. Magnetic properties of amorphous metallic alloys containing rare-earth and transition metal components, Physica, 91B (1977) 179-184

[55] A.K. Bhattacharjee, R. Jullien, M.J. Zuckermann, Magnetic properties of amorphous metallic alloys containing rare rarth impurities. J. Physics F, Metal Physics **7** (1977) 393-399

[56] Hui Zhang, Dechang Zeng and Zhongwu Liu, The law of approach to saturation in ferromagnets originating from the magnetocrystalline anisotropy, Journal of Magnetism and Magnetic Materials, 322 (2010) 2375-2380.

[57] J. Becker, A. Tsukamoto, A. Kirilyuk, J. C. Maan, Th. Rasing, P. C. M. Christianen, A. V. Kimel, Ultrafast magnetism of a ferrimagnet across the spin-flop transition in high magnetic fields, Physical Review Letters 118 (2017) 117203.

[58] R. C. Bhatt, Lin-Xiu Ye, N. T. Hai, Jong-Ching Wu, Te-Ho Wu, Spin-flop led peculiar behavior of temperature-dependent anomalous Hall effect in Hf/Gd-Fe-Co, Journal of Magnetism and Magnetic Materials, 537 (2021) 168196

[59] Dongdong Chen, Yaohan Xu, Shucheng Tong, Wenhui Zheng, Yiming Sun, Jun Lu, Na Lei, Dahai Wei, Jianhua Zhao Noncollinear spin state and unusual magnetoresistance in ferrimagnet Co-Gd, Physical Review Materials 6 (2022) 014402

[60] C. Fowley, K. Rode, Y-C Lau, N. Thiyagarajah, D. Betto et al. Magnetocrystalline anisotropy and exchange probed by high-field anomalous Hall effect in fully-compensated half-metallic $Mn_2Ru_xGa$ thin films (2019)

[61] Tianxun Huang, A Study of the behavior and mechanism of all-optical switching, Doctoral Thesis, University of Lorraine (2023)

[62] C. Banerjee, N. Teichert, K. E. Siewierska, Z. Gercsi, G.xY.P. Atcheson, P. Stamenov, K. Rode, J. M. D. Coey, J. Besbas, Single-pulse all-optical toggle switching of the

magnetization without Gadolinium in the ferrimagnet Mn$_2$RuGa, Nature Communications, 11 (2020) 4444

[63] R. Alben, J. J. Becker and M. C. Chi, Random anisotropy in amorphous ferromagnets, Journal of Applied Physics, 49 (1978) 1653-1658

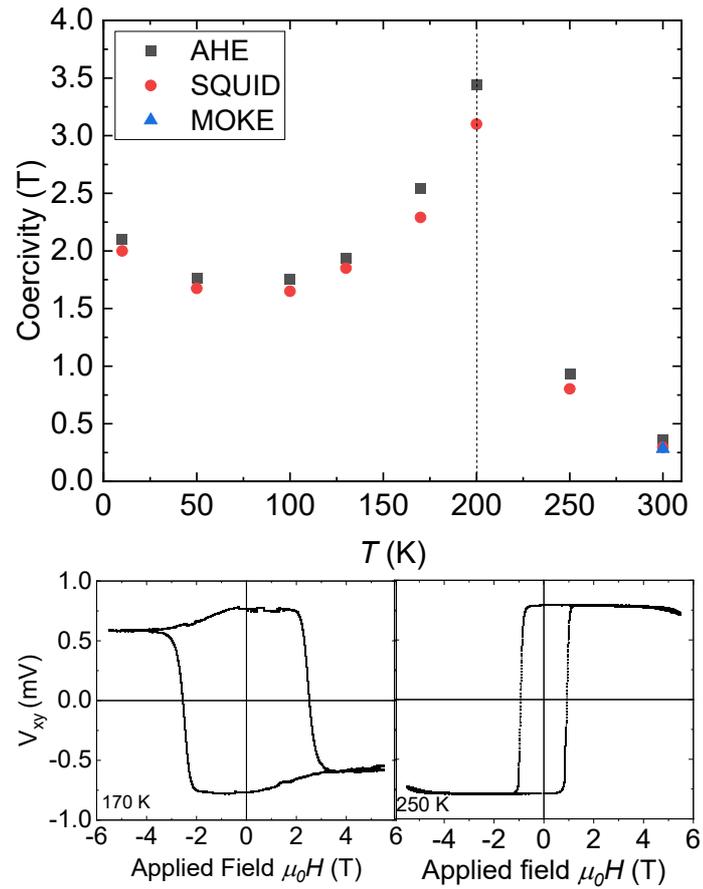

Figure S1. Coercivity for a-Tb$_{0.25}$Co$_{0.75}$ showing a divergence at the compensation temperature of 200 K. Data points are obtained by SQUID magnetometry or anomalous Hall effect. A magneto-optic Kerr effect point at 300 K is included. Include umbrella inserts Or OMIT as it is very similar to Fig 4